\documentclass[bib]{ba}

\usepackage{bayes}
\usepackage{amsmath,amsfonts,amsthm,amssymb,latexsym}
\usepackage{natbib}

\newcommand\dropcap\noindent

\begin{document}

\inserttype{article}
\author{Gelman, A. \&~Robert, C.P.}{%
  {\sc Andrew Gelman}\\{\em Department of Statistics and Department of Political Science, Columbia University}\\ 
  {\sf gelman@stat.columbia.edu}\\
  {\sc Christian P.~Robert}\\{\em Universit\'e Paris-Dauphine, CEREMADE, IUF, and CREST, Paris}\\{\sf
xian@ceremade.dauphine.fr}\\[.2cm]19 Oct 2012
}       
\title{The anti-Bayesian moment and its passing}

\maketitle

Over the years we have often felt frustration, both at smug Bayesians---in particular, those who object to checking of the fit of model to data, either because all Bayesian models are held to be subjective and thus unquestioned (an odd combination indeed, but that is the subject of another article)---and angry anti-Bayesians who, as we wrote in our article, strain on the gnat of the prior distribution while swallowing the camel that is the likelihood.  The present article arose from our memory of a particularly intemperate anti-Bayesian statement that appeared in Feller's beautiful and classic book on probability theory.  We felt that it was worth exploring the very extremeness of Feller's words, along with similar anti-Bayesian remarks by others, in order to better understand the background underlying controversies that still exist regarding the foundations of statistics.  We thank the four discussants of our article for their contributions to our understanding of these controversies as they have existed in the past and persist today.

\section*{1950:  the anti-Bayesian moment?}

Stephen Fienberg and Stephen Stigler broaden the historical perspective of our article and elaborate on our point that, as of 1950 (the year of publication of the first edition of Feller's classic probability text), the successes of Bayesian methods were not so well known, which is how Feller could get away with his disparaging attitude.

We agree with Stigler that much of the bringing of Bayesian methods to the mainstream of statistics came from publication of ``successful, clear, well-documented analyses and balanced discussion of choices made.''  What struck us about the Feller quote (and similar remarks from others over the years) was that he did not merely follow Thornton Fry and state the additional difficulties required to perform a Bayesian analysis; instead, Feller went all-in on an anti-Bayesian position.  It would have been easy enough for Feller to have patronizingly characterized Bayesian methods as interesting but unproven, but he felt the need to go further and actively warn his readers against the Bayesian approach.

Fienberg attributes Feller's attitude to a combination of personal quirks and a narrow, mathematically-focused intellectual environment.  This may be, but we speculate that two more factors were involved.  First, the neo-Bayesian movement associated with Jeffreys and Good had begun, an approach in which Bayesian inference was presented not merely as one of many approaches to statistics but as an overarching normatively necessary framework, thus, a potential threat to be fought.  Feller would not have felt the need to describe Bayesian arguments as ``dangerous'' unless there was some concern that scientists and engineers would fall to the siren song of Jeffreys (and this was indeed a possibility, as is indicated by the memoir of Jaynes which we quote in our article).

The second reason we suspect for Feller's rabidly anti-Bayesian stance is the postwar success of classical Neyman-Pearson ideas.  Many leading mathematicians and statisticians had worked on military problems during the Second World War, using available statistical tools to solve real problems in real time.  Serious applied work motivates the development of new methods and also builds a sense of confidence in the existing methods that have led to such success.  After some formalization and mathematical development of the immediate postwar period, it was natural to feel that, with a bit more research, the hypothesis testing framework could be adapted to solve any statistical problem.  In contrast, Thornton Fry could express his skepticism about Bayesian methods but could not so easily dismiss the entire enterprise, given that there was no comprehensive existing alternative.

If 1950 was the the anti-Bayesian moment, it was due to the successes of the Neyman-Pearson-Wald approach which was still young and growing, with its limitations not yet understood.  In the context of that comparison, there was no need for researchers in the mainstream of statistics to become aware of the scattered successes of Bayesian inference.  (In conversations with colleagues at Berkeley in the early 1990s, we were assured that hierarchical Bayesian models were nothing new:  David Brillinger referred to his work with Tukey in 1960, and Lucien LeCam referred to his applied Bayesian work in the late 1940s in France.  This difference was that, by the 1990s, the pure hypothesis testing approach was no longer central to the enterprise of theoretical statistics.  Thanks to the research of Stein, Efron, Morris, and others, hierarchical modeling ideas had entered the inner sanctum, and the distinction between classical and Bayesian approaches had blurred.)

\section*{Bayesian inference today}

As Deborah Mayo notes, the anti-Bayesian moment, if it ever existed, has long passed.  Influential non-Bayesian statisticians such as Cox and Efron are hardly anti-Bayesian, instead demonstrating both by their words and their practice a willingness to use full probability models as well as frequency evaluations in their methods, and purely Bayesian approaches have achieved footholds  in fields as diverse as linguistics, marketing, political science, and toxicology.  If there ever was a ``pure Bayesian moment,'' that too has passed with the advent of ``big data'' that for computational reasons can only be analyzed using approximate methods.  We have gone from an era in which prior distributions cannot be trusted, to an era in which full probability models serve in many cases as motivations for the development of data-analysis algorithms.  Meanwhile, the very success of applied Bayesian inference at the hands of Box, Rubin, and others has broadened our philosophy of Bayesian data analysis, and we now recognize model checking and expansion as key components of Bayesian data analysis, to be placed alongside the existing foundation of joint probability modeling of data and parameters.

According to Mayo, however, much of the field of philosophy is behind the times, still holding a simplified view of Bayesian inference as coherent reasoning in the spirit of Keynes, Neumann and Morgenstern, and other pioneers of the 1920s--1940s.  Similarly, in the popular and technical press, we have noticed that ``Bayesian'' is sometimes used as a catchall label for rational behavior.  We agree with Mayo that rationality (both in the common-sense and statistical meanings of the word) is complex.  At any given time, different statistical philosophies will be useful in solving different applied problems (see footnote 1 of \citealp{Gelman:2012}).  As Bayesian researchers, we take this not as a reason to give up in some areas but rather as a motivation to improve our methods:  if a non-Bayesian method works well, we want to understand how.

We do not see it as {\em necessary} that an engineering field such as statistics develop under multiple paradigms, but, given that this is what has happened, we seek to make the most of it.  Meanwhile, ideas develop, even in our short lifetimes.  For example, Mayo writes, ``While subjective Bayesians urge us to incorporate background information into the analysis of a given set of data by means of a prior probability \dots some of the most influential Bayesian methods in practice invite us to employ conventional priors that have the least influence on resulting inferences, letting the data dominate.''  Indeed, in the writing of the first two editions of {\em Bayesian Data Analysis} we felt uncomfortable with prior information and tried as much as possible to set up prior distributions in a purely structural way.  In recent years, however, various applied and theoretical experiences (similar to the reasons discussed by Wesley Johnson for avoiding noninformative priors) have moved us toward the use of stronger prior information in our inferences.  The gaps in practical Bayesian philosophy, as noted by Berger and others, can be seen as openings for methodological development.

As Johnson points out, as Bayesians we may be particularly accepting of multiple approaches to statistical reasoning, given that Bayesian reasoning itself proceeds by combining information (in Johnson's formulation, data and hypotheses from scientist teams A and B).  Bayesian inference as currently practiced (by ourselves and others) is not a good descriptive model of the scientific process, but we agree with Johnson that it is a useful point of comparison.  We also agree that a virtue of the Bayesian approach is it ``forces groups to lay their prior cards on the table.''  To be fair, a corresponding virtue of the classical statistical approach is its insistence on rigorous specification of data collection and decision rules.  To combine the words of Feller and Johnson, one might say that the Neyman-Pearson approach is particularly appropriate in settings with clean designs and sharp accept-reject decision rules, whereas the advantages of Bayes show themselves most clearly in problems of prediction.  This perhaps connects to Johnson's question of why the presence of Bayesian methods in the statistics curriculum seems to have lagged behind their use in practice.  To the extent that our teaching examples focus on the estimation of single parameters (the quest for $\theta$), the comparative advantages of Bayesian inference are less apparent.

\section*{Acknowledgements} 
This article is a rejoinder to the discussion of ```Not only defended but also
applied':  The perceived absurdity of Bayesian inference,'' for the {\em
American Statistician}. We thank the editor for organizing this discussion and
the commentators for their time and effort.  In addition, the first author (AG)
thanks the Institute of Education Sciences, Department of Energy, National
Science Foundation, and National Security Agency for partial support of this
work.  The second author's (CPR) research is partly supported by the French
Agence Nationale de la Recherche (ANR, 212, rue de Bercy 75012 Paris) through
the ANR 2011 BS01 010 01 projet Calibration.

\renewcommand{\bibsection}{\section*{References}}


\end{document}